\documentclass[12pt]{article}

\usepackage[utf8]{inputenc}
\usepackage[english]{babel}
\usepackage{color, graphicx}
\usepackage{amsmath}
\usepackage{graphicx}

\usepackage[dvipsnames]{xcolor}
\usepackage[pdftex,%
plainpages=false,%
linktocpage=true,%
a4paper=true,%
colorlinks=true,%
linkcolor=NavyBlue,%
citecolor=BrickRed,%
urlcolor=NavyBlue,%
bookmarks=true]{hyperref}

\newcommand{\CloudWatch}{{\sc CloudWatch}}
\newcommand{\CloudTrail}{{\sc CloudTrail}}
\newcommand{\GuardDuty}{{\sc GuardDuty}}
\newcommand{\Inspector}{{\sc Inspector}}
\newcommand{\WAF}{{\sc WAF}}
\newcommand{\Shield}{{\sc Shield}}
\newcommand{\Athena}{{\sc Athena}}
\newcommand{\Macie}{{\sc Macie}}
\newcommand{\KMS}{{\sc KMS}}
\newcommand{\Certificate}{{\sc Certificate}}
\newcommand{\AutoScaling}{{\sc AutoScaling}}
\newcommand{\LoadBalancing}{{\sc LoadBalancing}}
\newcommand{\CloudFront}{{\sc CloudFront}}
\newcommand{\SecurityGroups}{{\sc SecurityGroups}}
\newcommand{\Route}{{\sc Route53}}
\newcommand{\ElastiCache}{{\sc ElastiCache}}

\title{Cybersecurity in the AWS Cloud}
\author{Michael Soltys\footnote{California State University Channel
Islands, Professor in the Department of Computer Science,
URL: \href{http://www.msoltys.com}{www.msoltys.com}, Email:
\href{mailto:michael.soltys@csuci.edu}{michael.soltys@csuci.edu}}}
\date{\today}

\begin{document}
\maketitle

\begin{abstract}
This paper re-examines the content of a standard advanced course in
Cybersecurity from the perspective of Cloud Computing. More precisely,
we review the core concepts of Cybersecurity, as presented in a senior
undergraduate or graduate class, in light of the Amazon Web Services
(AWS) cloud.  
\end{abstract}

\section{Introduction}

This paper has three goals: (i)~to aid faculty in cloudifying a
Cybersecurity offering; (ii)~to re-examine Cybersecurity in light of
the new paradigm of Cloud Computing; and, (iii)~as a guide for
preparing for the AWS Security Specialty certification
(\cite{aws-security-cert}).  The paper presents an outline of
Cybersecurity, with topics examined in the context of AWS, and with a
long bibliography for a more in-depth study of each topic.  For a more
general guide to cloudifying a Computer Science curriculum
see~\cite{aws-soltys-cloudify}.

{\em Cybersecurity} is generally understood to be a set of techniques
and measures taken to protect digital information against unauthorized
access or attack.  Like all of Computer Science, it is a new field;
the first use of the term is recent: 1989.  Cybersecurity is also
known simply as {\em Security}, {\em Computer Security} and {\em
Information Security}. The prefix {\em cyber} comes from the word {\em
cybernetics}, which is the science of communication and control.
Cybernetics was imported into English in the 1940s, from the Greek
word: $\kappa\upsilon\beta\epsilon\rho\nu\eta\tau\eta\zeta$, {\em
kubernētēs}, which means {\em steersman}. Of course, kubernētēs now
also gave rise to {\em Kubernetes} \cite{aws-kubernetes}, an
open-source container coordination system.

It is difficult to give a precise definition to Cybersecurity, partly
because it has become such a vast field. Richard Bejtlich defines {\em
security} as {\em the process of maintaining an acceptable level of
perceived risk for a specified event} \cite{bejtlich-2019}.
Cybersecurity is the application of this principle to IT. There is no
{\em perfect security}, as Gene Spafford famously stated: {\em The
only truly secure system is one that is powered off, cast in a block
of concrete, and sealed in a lead-lined room with armed guards—and
even then, I have my doubts.} \cite{dewdney-1989}

Bruce Schneier makes the point regarding the security ROI that: {\em
Security is not an investment that provides a return, like a new
factory or a financial instrument.  It's an expense that, hopefully,
pays for itself in cost savings.  Security is about loss prevention,
not about earnings} \cite{schneier-2008}.
 
Schneier's quote points to a tension that practitioners experience in
the workplace: their advice regarding security expenditures is
frequently not heeded, and they are dismissed as prophets of ``doom
and gloom.'' When nothing bad happens they are forgotten, and when
breaches do occur they are blamed. Many decision makers seem to be
comfortable living with the possibility of a breach tomorrow rather
than spending today precious company resources (see
\cite{schneier-2019}).

AWS is conducive to the design of applications with security built in
from the beginning, not to mention a plethora of monitoring services
such as 
\CloudWatch,
\CloudTrail,
\GuardDuty,
\Inspector,
\WAF,
\Shield,
\Athena,
\Macie,
and others, disscussed in this paper.

\subsection{Approaches to Cybersecurity}\label{sec:approaches}

In this paper we are going to examine cybersecurity in the context of
Cloud Computing, but there are many other ways to focus on this vast
subject. In this section we list some of those other approaches.

{\bf As a Software Engineer:}
concentrating on how to write programs correctly and defensively,
e.g., avoid SQL injections or buffer overflows. Ed Amoroso (Networks
Security, AT{\&}T) writes:
\begin{quote}
{\em Software is most of the problem. We have to find a way to write
software which has many fewer errors and which is more secure.}
\cite[pg.~272]{cyberwar}
\end{quote}
There are two, unfortunately inadequate,  approaches to software
correctness: testing and formal methods; see
\cite{fred-taylor-2011}.
Typical development
techniques that have been shown to be effective are {\em code
minimization}, {\em formal development methods}\footnote{The problem
of ``how to write correct software'' has not been solved, and it is
one of the main open problems of Computer Science (see introduction
to~\cite{soltys-algs3}). Dave Parnas (\cite{Denning_2018}) is a
Software Engineer who has thought at length about this problem, and
has written on it extensively:
\cite{parnas-design,parnas-criteria,parnas-1983,softparns,parnas-risk}.
Parnas and Soltys co-authored the paper~\cite{parnas-soltys-fm06} on
the importance of mathematics for Software Design. There are
techniques to improve software quality, but no general methodology
exists as in more established areas of engineering. One thing
that Parnas promoted strongly throughout his career was the importance
of writing good documentation; one of the reasons the author became
fascinated by AWS was its culture of documenting its services; as
\cite{orban-2017} writes on page xxvii, {\em writing is deeply
ingrained in our [AWS] culture and decision-making process}.}, and
{\em using type-safe languages}. The point here is to make security a
``{\em built-in}'' rather than an ``{\em add-on}'';
see~\cite{Dykstra_2018}.

{\bf As an IT expert / system administrator:} install patches
and anti-malware applications, limit phishing attacks in your domain,
backups and system availability, etc.

{\bf As a cryptographer / cryptoanalyst:} design and analyze
cryptographic schemes (e.g., elliptic curve crypto) study issues of
implementation (e.g., OpenSSL libraries) and protocols (e.g., variants
of Kerberos).

{\bf As a business:} audits of compliance, risk assessment;
ultimately, Cybersecurity is a business decision, not an IT decision.
To see this note that IT could simply decide to encrypt everyone's
data with a secret key --- this would keep the data safe but useless
from the business perspective of the endeavor.  Here is a great quote
from AWS:
\begin{quote}
{\em
Security is the ability to protect information, systems and assets,
while delivering business value through risk assessment and threat
mitigation} \cite{aws-security-pillar-2018}.
\end{quote}

{\bf As an educator:} for example, teach basic practices for the
average user to be protected as much as possible.
It is important to educate the
public in the understanding that security and convenience are
orthogonal goals, in the sense that more security usually implies less
convenience; this can come as a surprise to many\footnote{See here for
typical event for small businesses \cite{small-business}.
Small businesses, which often cannot afford an IT
department, are especially vulnerable; 70\%\ go under following a
(successful) Ransomware attack.}: here is a quote from
the interesting article~\cite{botched-2018}:
\begin{quote}
{\em There is an inherent paradox to covert communications systems,
one of the former officials said: The easier a system is to use, the
less secure it is.}
\end{quote}

{\bf As a Cyber-warrior or law-enforcement:} defend
cyber-infrastructure, and probe and penetrate the cyber-infrastructure
of other countries (or organizations), all done within an organized
and legal framework. Or, in the area of {\bf digital forensics}, where
our Computer Science department at California State University Channel
Islands has a thriving partnership \cite{ci-httf}.

{\bf As a policy wonk:} which policies and regulations need to be in
place; what is the extent of the reach of law in digital forensics?
What international conventions ought to govern cyberwarfare?
See~\cite{cyberwar,stiennon-2015}.

{\bf As a cyber-vandal or criminal:} proving something that
everybody already knows: that destroying and breaking is always easier
than building and constructing.

\section{Cybersecurity core curriculum}

\subsection{Objectives of Cybersecurity}\label{sec:objectives}

In this section we list the classical objectives of Cybersecurity.
We briefly discuss the measures, techniques and procedures
that are usually deployed to meet those objectives.

\begin{enumerate}
\item {\bf Confidentiality:} in order to prevent the disclosure of
information to unauthorized entities (people or systems).  It is
usually achieved through data encryption, either with symmetric or
asymmetric (i.e., public key) algorithms. AWS has tools for managing
both keys and certificates: the Key Management Service\label{pg:kms}
(\KMS) manages keys for both symmetric and asymmetric encryption
\cite{aws-kms}. On the other hand, \Certificate\label{pg:certificate}
manages certificates \cite{aws-certificate} --- for example,
certificate for Secure Socket Layer (SSL) connections with DB
instances running the MariaDB engine (\cite{aws-mariadb-ssl}).  

\item {\bf Integrity:} data cannot be modified undetectably.  An
example where data integrity is essential is in inter-bank money
transfers; changing the amount in transit would be even more damaging
to the banking system than disclosing the amount --- of course, we can
have both integrity {\em and} confidentiality.  Another measure for
integrity is {\em digital signatures} (also known as {\em digital
digests}) which both authenticate a document and ensure its integrity.
This is achieved with hashing functions and public key cryptography.
``Adobe Sign'' (which, as all of Adobe, is built on AWS
\cite{adobe-aws}) is an example
of a technology implementing digital signatures.

\item {\bf Availability:} information is available when it is needed,
that is, both the computer system, and the communication channels are
functioning correctly.  A typical attack against availability is the
{\em (distributed) denial-of-service} attack (DDoS) \cite{aws-ddos},
but of course not all attacks against availability are malicious;
Werner Vogel\footnote{Werner Vogel is Vice President \&\ Chief
Technology Officer at Amazon.com}:
{\em Everything fails all the time.}

Best practice for availability is to design {\em fault tolerant} and
{\em loosely coupled} systems.  AWS provides mechanisms for designing
such system with, for example, \AutoScaling\label{pg:autoscaling} 
(\cite{aws-as}) which is a service that, when deployed, grows and
shrinks the number of EC2 instances assigned to a task according to
demand, and
\LoadBalancing\label{pg:loadbalancing} 
(\cite{aws-elb}) which is a service that distributes tasks among a
fleet of EC2 instances.  Two other important concepts associated with
availability are {\em Multi Availability Zone Deployment}, and Content
Distribution Networks (CDN) which are implemented at {\em edge
locations} with \CloudFront\label{pg:cloudfront}, which is essentially
a giant caching service.
\cite{aws-cloudfront}. 
\end{enumerate}

We abbreviate the foundational triad of
Confidentiality, Integrity and Availability as CIA.  
The CIA objectives are often supplemented by the following ``three
A's'' objectives:
\begin{enumerate}
\setcounter{enumi}{3}
\item {\bf Authentication:} to ensure that data, transactions,
communications or documents are genuine; a typical example of
authentication is a login/password pair, often supplemented with {\em
Multi-Factor Authentication} (MFA), to verify the user of a system.
An authentication scheme purports to validate that the parties
involved are who they claim to be, but technically only verifies that
the agent attempting to gain permission is in possession of the
credentials --- this could be by legitimate ownership, or by stealing
them, guessing them, or generating them by a brute-force search if the
credentials were short strings or dictionary words. 

AWS has a powerful service for authentication: Identity Access
Management (IAM) (\cite{aws-iam}), which coordinates access through:
{\em users}, {\em groups}, {\em roles} and {\em policies}.  AWS also
provides Active Directory (\cite{aws-active-dir}) and SAML
(\cite{aws-saml}) for authentication.

\item {\bf Authorization:} once an agent is authenticated, where the
`agent' can be a human user or a machine process, authorization
stipulates what this agent is allowed to do. This is usually done
through a {\em policy}, e.g., an AWS S3 policy which stipulates
whether the agent has read access to a given S3 bucket 
(\cite{aws-gist-s3-policy}). The
guiding principle of authorization is:
\begin{quote}
the {\em Principle of Least Privilege}: give every agent the minimal
amount of permissions that are required to get the job done.
\end{quote}
Fans of spy thrillers will recognize this as the ``need to know''
principle. In AWS, both authentication and authorization are
implemented with the {\em Identity Access Management} (IAM) service
discussed in ``Authentication'' above.

\item {\bf Accounting:} Keeping record logs, and automating parsing
them and reacting in near real-time.  Related AWS services are as
follows:
\begin{enumerate}
\item \CloudTrail\label{pg:cloudtrail} 
(\cite{aws-security-scale-2015}), which keeps track
of all API calls, and logs them in an S3 bucket; 

\item \CloudWatch\label{pg:cloudwatch} (\cite{aws-cloudwatch}), which
monitors service usage, and is perhaps one of the most used tools in
the AWS arsenal;
\begin{figure}[h]
\begin{center}
\includegraphics[width=12cm]{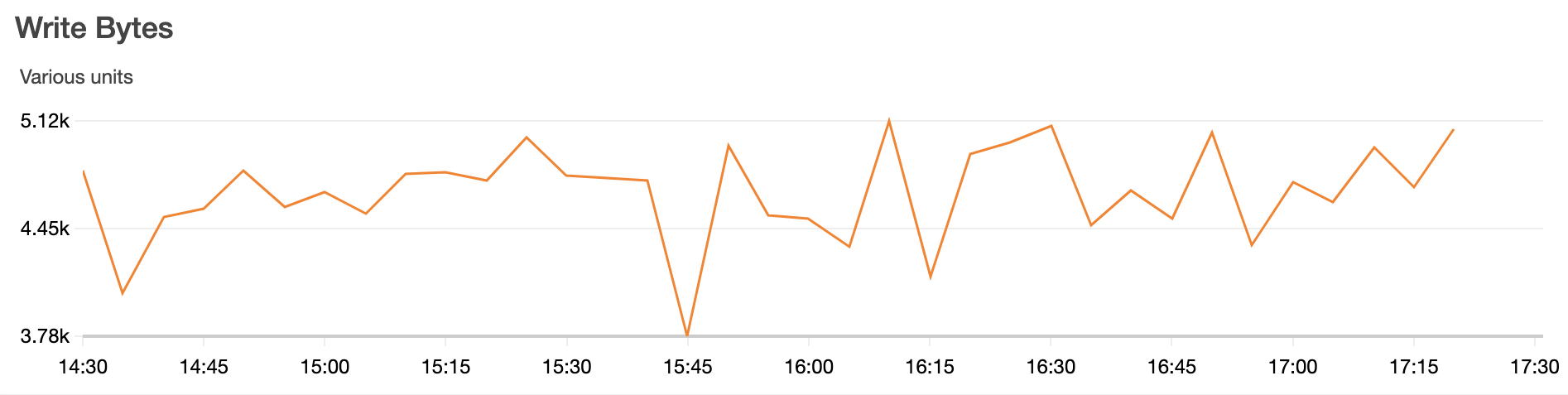}
\end{center}
\caption{A \CloudWatch\ metric that shows bytes written to an Elastic
Block Store (EBS) over a period of 3 hours.}
\end{figure}

\item \GuardDuty\label{pg:guardduty} (\cite{aws-guardduty}), which is
a continuous monitoring service that analyzes and processes the
following data sources: Virtual Private Cloud (VPC) flow logs,
\CloudTrail\ and DNS logs.

\item \Inspector\label{pg:inspector} (\cite{aws-inspector}), which
tests the network accessibility of EC2 instances and the security
state of applications that run on those instances. Inspector assesses
applications for exposure, vulnerabilities, and deviations from best
practices. After performing an assessment, Inspector produces a
detailed list of security findings that is organized by level of
severity.

\item \Macie\label{pg:macie} (\cite{aws-macie}) and
\Athena\label{pg:athena} (\cite{aws-athena}) which uses machine
learning to discover and classify sensitive data in S3, and an
interactive query service that makes it easy to analyze data directly
in S3 using SQL queries, respectively.
\end{enumerate}
\end{enumerate}

Finally, to CIA $+$ AAA we add:

\begin{enumerate}
\setcounter{enumi}{6}
\item {\bf Non-repudiation:} implies that one party of a transaction
cannot deny having received a transaction nor can the other party deny
having sent a transaction; an example of application would be online
bidding. Note that non-repudiation can also be seen under Accounting.
\end{enumerate}
We achieve some of the eight objectives through cryptography, but not
all. For example, \SecurityGroups\label{pg:securitygroups} (outside
the cloud known as {\em firewalls}), filter network traffic
(\cite{aws-security}), and \LoadBalancing, which help in the scaling
of demand and discussed in the section on ``Availability,'' are not
cryptographic applications.

\subsection{Examples of basic attacks}

\subsubsection{Phishing}

{\em Phishing}\index{Phishing} is a social engineering attack that
aims at the weakest link of any computer system: the human user.
Phishing attacks have become extremely sophisticated over the years.
They range from SPAM emails, to targeted attacks (e.g., {\em
whaling}, which are emails aimed at high profile targets such as CFOs
of a company) where the perpetrators study their victim, sometimes for
months, before crafting a precisely targeted phishing email.  See for
example~\cite{phishing}.

Although not technically phishing, there are many varieties of online
fraud that are similar to phishing. For example {\em trolling}
\cite{chen-2015}, where users --- often state actors --- post inflammatory
and digressive material, {\em identity theft}, {\em cyberbullying},
{\em evil twin} attacks where users --- again, often state actors ---
impersonate people and companies, frequently for financial gain or
defamation. Online fraud is big business: \cite{criminals-cctv}.

\subsubsection{DDoS}

Distributed Denial of Service, DDoS, was briefly reviewed in
Section~\ref{sec:objectives}, under ``Availability,'' where we
mentioned the excellent AWS whitepaper on the subject \cite{aws-ddos}.
AWS customers have three tools at their disposal for DDoS {\em
mitigation}: 
\begin{enumerate}
\item \Route\label{pg:route}, which is a DNS service \cite{aws-route53}.
\item \CloudFront, which is a Content Distribution Network, discussed
in Section~\ref{sec:objectives} under ``Availability.''
\item \Shield\label{pg:shield}, which protects against DDoS attacks,
but costs \$3K/month, and so it is more
of an enterprise solution than a private customer solution. See the
AWS whitepaper on DDoS resiliency (\cite{aws-ddos}) and the AWS
developer guide on Shield itself (\cite{aws-shield}).
\end{enumerate}
Note that all three tools work with {\em Edge Locations}, which are
physical data centers located in key cities; edge locations work as
giant caches\footnote{The Amazon Builder's Library
(\cite{aws-builders-lib}) is a magnificent source of information on
the design of AWS services; information that is well beyond the scope
of the AWS Security certificaton. But, reading the writeups
from the Builder's Library 
it becomes apparent that {\em latency} is the
biggest issue in the Cloud, and that {\em caching} is the prevalent solution
to that problem. Caching has a beautiful algorithmic theory, based on
{\em Online Algorithms}, that is worth knowing to have more insight
into the AWS cloud. For example, the reader is directed to
\S{5.2}, {\em Paging}, in~\cite{soltys-algs3}.}. AWS has an actual 
caching tool
for accelerating applications, \ElastiCache\label{pg:elasticache} 
\cite{aws-elasticache}. It
supports two open-sourced in-memory caching engines: Memcached and
Redis.

An advantage of edge locations is that threats can be mitigated there
rather than on the server hosting a particular application. This puts,
as it were,
distance between the problem and the application.

\subsubsection{SQL injection}\label{sec:sql}

The Open Web Application Security Project (OWASP) \cite{owasp} is a
nonprofit foundation that works to improve the security of software.
OWASP publishes a ranking of the 10 most-critical web application
security flaws, which are known as the ``OWASP Top 10.'' It is not
surprising that ``injection attacks'' are currently in the number one
spot on that list. 

AWS's WAF\label{pg:waf}, Web Application Firewall
(\cite{aws-waf-owasp}), helps to defend a website against the attacks
listed by OWASP.  It is important to note that WAF does not help
mitigating DDoS; \Shield, discussed in the previous section, along
with \CloudFront\ and \Route, are used to defend against DDoS.

\subsubsection{Malware}\label{sec:malware}

In the popular imagination, cybersecurity {\em is} about malware, in
its various forms: {\em Viruses}, which modify legitimate host files
in such a way that when the file is executed, the virus is also
executed.  {\em Worms}, which self-replicate across a network, without
end-user action (unlike Viruses, which require that an end-user
``click on it''). {\em Trojans}, which have replaced worms and which
masquerade as legitimate programs. {\em Ransomware}, which implements
digital extortion by demanding a ransom pay for decrypting the user's
files. 

Malware is frequently imported into a user's computer by a phishing
scam, and sophisticated instances deploy {\em zero-day exploits} in
order to escalate privileges in the Operating System. A zero-day
exploit is a software vulnerability for which no patch has been
released yet (zero days since the release of a patch). Zero-day
exploits can still be effective after the patch has been released if
it has not been installed by the system administrator.  The
vulnerability exploited by the zero-day can be used to gain privileges
in the Operating System.  

There is also {\em Adware} and {\em Spyware}, with eponymous
functionality.

{\bf Case Study -- Stuxnet:}
See~\cite{zetter-2014,sanger-2018,Kirkpatrick_2019}.  Stuxnet was a
sophisticated worm, developed by the US and Israeli governments
(\cite{sanger-2012}) around 2005, code name ``Olympic Games'', aimed
at the uranium enrichment facility in Natanz (Iran), and discovered by
the Infosec community around 2010. The worm attacked {\em Programmable
Logic Controllers} (PLC) manufactured by the German company Siemens,
used at the Natanz facility to run centrifuges.  Stuxnet is
considered to be the first deployment of a cyberweapon agains IoT.
AWS has a comprehensive offering in the security of IoT
\cite{aws-iot}.

{\bf Case Study -- Heartbleed:}
{\em Heartbeat} is an extension protocol for the {\em Transport Layer
Security} (TLS), which works as an acknowledgment mechanism, which is
a way to verify that a server is up, i.e., whether it has a
``heartbeat''. The protocol sends a 40Kb message, and asks for it to
be repeated back; the receiver of the message allocates a memory
buffer, stores the data, reads the data, and sends it back. The
OpenSSL library was shipped with heartbeat support enabled (March
2012). There was a buffer overflow vulnerability in the protocol that
was not discovered until April 2014.

{\bf Case Study -- Olympic Destroyer:}
For this case study see~\cite{olympic-destroyer}. The Olympic
Destroyer was a worm targetted at the Pyeongchang, South Korea, Winter
Olymbic games of 2018. The well crafted worm brought down the IT
infrastructure of the games.  This case study is a great example of
the problem of {\em attribution}, i.e., where does the attack
originate and who is responsible.

{\bf Case Study -- Sandworm:}
See~\cite{greenberg-2019}.  Sandworm is
not malware per se, but rather a group of state-sponsored hackers,
according to experts. For example, they are believed to have planted
malware inside the US electric utilities in 2014.  The citation for
this case study is a book which relates malware to geopolitical
issues. It also reads as a great thriller, with vivid descriptions of
the people involved in the drama, as well as cryptic references to the
movie {\em Dune}.

Finally, in order to understand the potential for mischief, it is good
to start by reading the seminal paper~\cite{thompson-1984}.  A free
malware tool can be accessed at \cite{virus-total}.

\subsection{Cryptography}

Shafi Goldwasser (MIT) defines Cryptography as {\em the art of {\em
computing} and {\em communicating} in the presence of an {\em
adversary}} \cite{goldwasser-2008}, and Oded Goldreich writes that
{\em Cryptography is concerned with the conceptualization, definition,
and construction of computing systems that address security concerns}
\cite{goldreich1}. For the sake of this paper, Cryptography is a set
of mathematical tools, which, when implemented as computer programs,
help us achieve some of the security objectives listed in
Section~\ref{sec:objectives}.

While AWS Security Specialty certification does not require an
in-depth understanding of {\em cryptographic protocols} (aka, {\em
cryptoschemes} and {\em cryptosystems}), they are nevertheless
foundational and required for anyone who wants to work in the field of
security.  

\subsubsection{Basic concepts}

The three common services of cryptography are the following:
\begin{enumerate}
\item {\bf Encryption/Decryption:} basic service of cryptography,
that enables the sending of data between participants in a way that
prevents others from reading it: \\
\begin{tabular}{ccccc}
plaintext & $\stackrel{encryption}{\longrightarrow}$
& ciphertext & $\stackrel{decryption}{\longrightarrow}$
& plaintext \\
{\tt ATTACKATDAWN} & $\stackrel{encryption}{\longrightarrow}$
& {\tt HAAHJRHAKHDU} & $\stackrel{decryption}{\longrightarrow}$
& {\tt ATTACKATDAWN}
\end{tabular}
where the example uses the Caesar cipher with key $k=7$.

\item {\bf Integrity checking:} reassuring the recipient of a message
that the message has not been altered since it was generated by a
(legitimate) source.

\item {\bf Authentication:} verifying someone's (something's)
identity; i.e., making sure that the source is legitimate and is who
they claim to be.
\end{enumerate}
Note the parallel to CIA and AAA listed in
Section~\ref{sec:objectives}: clearly encryption/decryption serve
confidentiality; integrity checking obviously ensures integrity; and,
authentication is the first `A' in `AAA.'

Cryptographers invent protocols; cryptoanalysts attempt to break them.
A cryptographic system consists of an algorithm and a secret value,
aka, a {\em key}. It is like a combination lock; everybody knows how
it works, but you will not open it without a key. The security of a
cryptoscheme depends on how much work a ``bad guy'' needs to do to
break it.

We consider a cryptoscheme {\em secure} if there does not exist a way
for finding a key that is substantially better than a brute force
search for a working key. With that in mind, the following are all
equivalent: (i)~it is possible to do secure sessions; (ii)~there exist
pseudo-random generators; (iii)~there exist ``one-way functions'';
(iv)~there exist secure digital signature schemes.
This equivalence is known as the {\em Fundamental Theorem} of
Cryptography \cite{rackoff-lectures}.  
Note that we have excellent candidates for each of the
above, but we lack a proof in each case! This is a fundamental gap in
the scientific understanding of cryptography, and a fundamental open
problem of Computer Science.
The above also suggests that in practice a good source of {\em
pseudo-randomness} is necessary in order to achieve security. In
particular, the key ought to be randomly chosen.

In light of the lack of a proof that any particular cryptosystem
(except the {\em one-time pad}) is secure, we use the following
working definition of security: lots of smart people have been trying
to figure out how to break $X$, but so far they have not been able to
come up with anything yet.  Therefore $X$ is considered ``secure''.
This is known as the {\em Fundamental Tenet} of cryptography.

Finally, the {\em Fundamental Assumption} of cryptography is that
security does not rely on obscurity (where it often does in
Cybersecurity). What this means is that it is always assumed that
everyone knows how a particular cryptoscheme works (i.e., it is ``open
source''), that is the algorithm is public knowledge. The secret is
the key.  So in principle any cryptoscheme can always be broken, by,
say, brute-force search, but in practice it is too much work for the
``bad guy.''
There are three classical attacks against cryptoschemes, that is,
approaches to finding out the plaintext and key:
\begin{enumerate}
\item {\bf Ciphertext only:}  The attacker has only the ciphertext,
and has to compute the plaintext and key from it.

\item {\bf Known plaintext:}  The attacker has both the ciphertext and
the corresponding plaintext, and has to compute the key from it.

\item {\bf Chosen plaintext:}  The attacker can choose any plaintext,
and get the corresponding ciphertext, and has to compute the key from
it.
\end{enumerate}

A good cryptosystem should resist all 3 attacks.

\subsection{Symmetric encryption}

In symmetric encryption the same {\em secret key} is used for both 
encryption and
decryption. A widespread symmetric function is the {\em Advanced
Encryption Standard} (AES) cryptographic scheme; AES with secret 
keys of size
256 is the de facto standard for
symmetric encryption at AWS \cite{aws-crypto}:
$f_{\text{AES}}:X\times Y\longrightarrow Z$, where $X\in\{0,1\}^k$ is
the key where $k\in\{128,192,256\}$, and $Y,Z\in\{0,1\}^{128}$, where
$Y$ is the {\em plaintext} block and $Z$ is the {\em ciphertext}
block.
See \S{3.5}
in~\cite{kaufman} for an excellent description of $f_{\text{AES}}$. 

It is important to keep in mind that there are two equally important
layers in cryptography, both in symmetric and asymmetric. The first
layer is the mathematical presentation of a cryptographic function,
such as $f_{\text{AES}}$.
The second layer is the programmatic implementation of
$f_{\text{AES}}$. For example, OpenSSL (\cite{openssl}) implements the
most common cryptographic functions (type \verb|openssl list-cipher-commands| 
to see which). We can encrypt a file
\verb|plaintext| with OpenSSL using AES as follows:
\begin{verbatim}
openssl enc -aes-256-cbc -pass pass:password -p -in plaintext
\end{verbatim}
Note that \verb|-aes-256-cbc| means that we used AES with a key of
size 256 (largest possible), and we used {\em Cipher Block Chaining}
(CBC) (\S{4.2.2} in~\cite{kaufman}) to encrypt files that have more
than 128 bits. The \verb|-p| switch means that the {\em salt} and the
{\em initial vector} (iv) and the key resulting from the
\verb|password| are output as well.  The salt and iv are not part of
$f_{\text{AES}}$, but they are part of the implementation in order to
guard against ``batch attacks'' (see pg.~243 in~\cite{kaufman}).  The
salt in the implementation illustrates well the difference between the
mathematical function $f_{\text{AES}}$ and its implementation in
OpenSSL.  From the point of view of practitioners of Cybersecurity,
the implementation is at least as important as the mathematics
defining $f_{\text{AES}}$. 

As mentioned in Section~\ref{sec:malware}, it was an implementation
vulnerability in OpenSSL that lead to Heartbleed.  Thus, it is
important not to confuse the mathematical strength, however defined,
of a cryptographic function with the security offered by the
implementation of that function. The implementation may be problematic
(we come back to the issue of buggy software described in
Section~\ref{sec:approaches}), or the implementation may be correct
but the deployment of the application may be faulty. 

The point that we belabored here is that cryptographic benefits do not
translate automatically to security benefits. 

\subsection{Asymmetric, aka Public Key, encryption}

\newcommand{\priv}{k_{\mathrm{priv}}}
\newcommand{\pub}{k_{\mathrm{pub}}}

A {\em Public Key Cryptosystem} (PKC) consists of three sets: $K$, the
set of pairs of {\em keys}, $M$, the set of {\em plaintext} messages,
and $C$, the set of {\em ciphertext} messages.  A pair of keys in $K$
is $k=(\priv,\pub)$; the {\em private} key and the
{\em public} key, respectively.  For each $\pub$ there is a
corresponding {\em encryption} function $e_{\pub}:M\longrightarrow C$
and for each $\priv$ there is a corresponding {\em decryption}
function $d_{\priv}:C\longrightarrow M$.

The property that the encryption and decryption functions must satisfy
is that if $k=(\priv,\pub)\in K$, then $d_{\priv}(e_{\pub}(m))=m$ for
all $m\in M$.  The necessary assumption is that it must be difficult
to compute $d_{\priv}(c)$ just from knowing $\pub$ and $c$.  But, with
the additional {\em trapdoor} information $\priv$, it becomes easy to
compute $d_{\priv}(c)$.

The three classical PKCs are: Diffie-Hellman, which is not really a
PKC but rather a way of agreeing on a secret key over an insecure
channel, as well as ElGamal and RSA.  All three require large primes
(in practice at least~2,000 bit long); a single prime for
Diffie-Hellman and ElGamal, and a pair of primes for RSA.  Those large
primes are usually computed using the Rabin-Miller algorithm --- see
\S{6.4} in~\cite{soltys-algs3}.

An example of PKC in AWS is the generation of a key pair when
launching an EC2 instance \cite{aws-ec2-ug}. The keys that EC2 uses
are 2048-bit SSH-2 RSA keys. 

\section{AWS best practices}

The AWS {\em Well-Architected Framework} \cite{aws-well-arch-2019}
proposes {\em five pillars} for the design of cloud infrastructure: 
\begin{enumerate}
\item Operational Excellence
\item {\bf Security}
\item Reliability
\item Performance Efficiency
\item Cost Optimization
\end{enumerate}
We are going to focus on the security pillar
\cite{aws-security-pillar-2018}. 

Although in theory an ``on-premises solution'' can achieve the same
level of security as a ``cloud solution,'' in practice bespoke
security solutions suffer from three common shortcomings: 
\begin{enumerate}
\item the preponderance on manual processes, rather than automated
solutions, e.g., IT visually inspects logs at the end of the day; 
\item eggshell security models, where the defense is at the perimeter,
and once that is breached, e.g., a password is stolen, the intruders
have the keys to the realm; 
\item insufficient auditing, e.g., are those
logs really examined, and what is being logged in the first place? 
\end{enumerate}
On the other hand, the homogeneity of the cloud, e.g., all API calls
are recorded by AWS \CloudTrail\ (\cite{aws-security-scale-2015}), and
the existence of tools to automate response, allow for much better
security\footnote{The ``homogeneity of the cloud'' may be a
double-edged sword: yes, it is easier to prepare defenses for attacks
on a uniform platform, but attackers can also take advantage of the
uniformity in deploying the same attacks against multiple targets,
where they would have to adjust their approach if the targets were on
very different platforms. Here is an interesting question; can the
cloud infrastructure be ``salted'' in some sense to avoid batch attacks?}.  
AWS proposes the following design
principles to strengthen security in the cloud (see page~2
of~\cite{aws-security-pillar-2018}):
\begin{enumerate}
\item Implement a strong identity foundation, using the principle of
{\em least privilege} (discussed in Section~\ref{sec:objectives}),
enforce separation of duties, and require appropriate authorization
for all interactions with AWS resources.
\item Enable traceability, with monitoring, alert and audit actions,
and change to the environment in real time. Here is where the strength
of the cloud comes to the fore. Also, integrate logs and metric with
systems to automatically respond and take action.
\item Apply security to all layers, rather than security only at the
outer layer; that is, apply what is called {\em defense-in-depth} with
other security controls.
\item Automate security best practices, so that with software-based
security mechanism it is possible to securely scale more rapidly and
cost effectively. Implement controls that are defined and managed as
code in version controlled templates.
\item Protect data in transit and at rest.
\item Enforce the principle of least privilege, by giving access to
data only to those agents who really need the access. An approach to
implementing this is to start by denying access to everything and
allowing access as the need arises.
\item Prepare for security events, by having an incident management
process aligned to the organizational requirements.
\end{enumerate}
The above should be seen in the context of the AWS {\em Shared
Responsibility Model} (\cite{aws-shared-responsibility}), where AWS is
responsible for protecting its global infrastructure (security {\em
of} the cloud), and the customers are responsible for securing the
resources that they create (security {\em in} the cloud).  All the AWS
compliance programs can be seen in \cite{aws-compliance}.

\section{Acknowledgments}

We are grateful to Sami Al-Salman, Sam Decanio and Eric Gentry 
(California State University Channel
Islands) for helpful comments regarding an early draft.

\section{Appendix}

This section contains a summary of all the AWS services mentioned in
this paper.

\bigskip

\noindent
\begin{tabular}{|l|l|c|c|}\hline
{\bf AWS Service} & {\bf Short description} & {\bf Pg} & {\bf Cite} 
\\\hline\hline
\Athena & S3 data analysis with SQL queries
& \pageref{pg:athena} & \cite{aws-athena} \\
\AutoScaling & Grows and shrinks the number of EC2s
& \pageref{pg:autoscaling} & \cite{aws-as} \\
\Certificate & Manages SSL/TLS certificates 
& \pageref{pg:certificate} & \cite{aws-certificate} \\
\CloudFront & A caching mechanism at edge locations
& \pageref{pg:cloudfront} & \cite{aws-cloudfront} \\
\CloudTrail & Keeps track fo all API calls
& \pageref{pg:cloudtrail} & \cite{aws-security-scale-2015} \\
\CloudWatch & Monitors services usage 
& \pageref{pg:cloudwatch} & \cite{aws-cloudwatch} \\
\ElastiCache & A caching service
& \pageref{pg:elasticache} & \cite{aws-elasticache} \\
\GuardDuty & Monitors logs of VPC, \CloudTrail, DNS
& \pageref{pg:guardduty} & \cite{aws-guardduty} \\
\Inspector & Network accessibility of EC2
& \pageref{pg:inspector} & \cite{aws-inspector} \\
\KMS & Manages encryption keys 
& \pageref{pg:kms} & \cite{aws-kms} \\
\LoadBalancing & Distributes tasks among EC2s
& \pageref{pg:loadbalancing} & \cite{aws-elb} \\
\Macie & Machine Learning discovery of sensitive data in S3
& \pageref{pg:macie} & \cite{aws-macie} \\
\Route & DNS service
& \pageref{pg:route} & \cite{aws-route53} \\
\SecurityGroups & Virtual firewall at the EC2 layer
& \pageref{pg:securitygroups} & \cite{aws-security} \\
\Shield & Protects agains DDoS attacks
& \pageref{pg:shield} & \cite{aws-shield} \\
\WAF & A web application firewall
& \pageref{pg:waf} & \cite{aws-waf-owasp} \\
\hline
\end{tabular}

\nocite{*}
\bibliographystyle{IEEEtran}
\bibliography{soltys_cloudsec}

\end{document}